# Probing topological invariants in the bulk of a non-Hermitian optical system

Julia M. Zeuner[1], Mikael C. Rechtsman[2], Yonatan Plotnik[2], Yaakov Lumer[2], Mark S. Rudner[3], Mordechai Segev[2], and Alexander Szameit[1]

[1]*Institute of Applied Physics, Abbe Center of Photonics, Friedrich-Schiller-Universität Jena, Max-Wien-Platz 1, 07743 Jena, Germany*

[2]*Department of Physics, Technion – Israel Institute of Technology, Haifa 32000, Israel*

[3]*The Niels Bohr International Academy, Niels Bohr Institute, 2100 Copenhagen, Denmark*

**Abstract**

**Topological insulators are insulating in the bulk but feature conducting states on their surfaces. Standard methods for probing their topological properties largely involve probing the surface, even though topological invariants are defined via the bulk band structure. Here, we utilize non-hermiticy to experimentally demonstrate a topological transition in an optical system, using bulk behavior only, without recourse to surface properties. This concept is relevant for a wide range of systems beyond optics, where the surface physics is difficult to probe.**

The notion of topological protection of electronic properties was first explored by Thouless and coworkers [1], who demonstrated that the Hall conductance of a two-dimensional electron gas (with Fermi energy placed within a bulk gap) is proportional to an integer-valued topological quantity. The global nature of the topological quantity introduces a striking robustness: small changes to the system (including the addition of disorder) have almost no effect on the Hall conductance [2]. Non-trivial topology implies the presence of conducting surface states (via the "bulk-edge correspondence principle"), which play a central role in many topological phenomena. A subsequent resurgence of interest in topological phenomena began with the prediction and observation of the quantum spin Hall effect [3–5], followed by the prediction [6,7] and observation [8] of an analogue for microwave photons. This scheme relies on the large magnetic response occurring in the microwave regime, which is not generalizable to optical photons. Many theoretical proposals were subsequently put forward on how to achieve topologically protected optical transport [9–12] and only recently it was experimentally realized for the first time using photonic waveguide arrays [13]. A different structure, relying on a two-dimensional network of coupled resonators, was independently used to implement this phenomenon in a silicon platform [14]. Relatedly, one-dimensional quasicrystals were demonstrated to have a direct mapping to two-dimensional topological physics [15]; topological modes were observed in quantum walks [16]; as well as topological modes in radiating systems [17]. Finally, recent progress demonstrated topological phenomena in ultracold gases, such as the direct measurement of the Zak phase [18]. However, in typical cold atoms experiments, accessing edge physics is extremely challenging due to the smooth quadratic trap holding the atoms. Accordingly, measuring topological properties in such systems must be carried out by a complex procedure of sweeping the wavefunction back and forth through the Brillouin zone [19–22]. Similar ideas have been proposed for exciton-polariton condensate systems [23].

Concurrently, notions of topological physics were introduced also in essentially non-Hermitian systems [24,25], where dissipation plays a central role. Under appropriate conditions, the evolution history of the wavefunction in a non-Hermitian system may encode topological features of the system's Bloch bands, which are reflected in robustly quantized values of certain time-integrated observables [24]. This discovery is important first because it identifies a methodology for unraveling the topological behavior of non-Hermitian systems. Second, it proposes a simple way to study the topological features of the physical system by examining the history of the bulk wavefunction, something that would otherwise require complicated protocols [18,22] or probing edge physics [5,13,26], which is difficult for many systems. However, [24] studied a discrete lattice model. In *continuous* periodic systems, such as solid state or photonic structures and superconducting circuits, deviations from the idealized tight-binding description can be significant. This is especially significant in non-Hermitian systems, where loss is added into the model phenomenologically, in reality representing coupling to radiating states. This brings up many interesting questions. In particular, to what extent can the conceptual idea suggested in [24] be applied to physical systems with continuous degrees of freedom? Can the topological features of such a non-Hermitian system be revealed by bulk experiments alone?

Here, we do just that: we demonstrate the first experiment to probe the topological properties of a non-Hermitian (effectively quantum) system. We show that its topological features can be unraveled by bulk experiments. Specifically, we study an optical setup designed to approximately realize the idealized one-dimensional non-Hermitian lattice model of [24] termed the non-Hermitian dimer model (NHDM)). We show that this conceptual idea can be implemented in a true non-Hermitian physical system, even though the physics of the system is considerably more involved.

The NHDM resembles the Su-Schrieffer-Heeger (SSH) model of polyacetylene [27], also called the dimer model, defined in terms of hopping on a one-dimensional lattice with alternating strong and weak bonds. The SSH model features topologically distinct phases, corresponding in the simplest case to two inequivalent dimerization patterns. The NHDM contains one additional ingredient, which is that particles can be lost, or absorbed, at every second site; this is the non-Hermitian component.

We implement the NHDM in a photonic setting, although the concept is completely general. Our system is composed of an array of evanescently-coupled waveguides, depicted schematically in Fig. 1(a). A sketch of the input facet of the array is shown in Fig. 1(b) labeling the different sublattices A and B and unit cell index, m. The presence of edge states in a photonic dimer waveguide array has been demonstrated experimentally [28]. The 'strong bonds' of the SSH model correspond to waveguides that are close to one another, and the 'weak bonds' correspond to the larger spacing. The length of the unit cell is $d$, and the spacings between A and B sublattices are $d_1$ and $d_2$. The effective non-hermiticity is introduced by oscillating every second waveguide sinusoidally in the propagation direction, which introduces radiation loss. Such a spatially-oscillating waveguide mimics a temporally-oscillating potential well in a quantum mechanical system, where the oscillation introduces leakage to high energy unbound continuum modes [29].

The full continuum description of the propagation of light through such a structure is the paraxial approximation to Maxwell's equations, which is mathematically equivalent to the Schrodinger equation:

$$i\partial_z \psi(x,y,z) = -\frac{1}{2k_0}\nabla^2 \psi(x,y,z) - \frac{k_0 \Delta n(x,y,z)}{n_0}\psi(x,y,z), \qquad (1)$$

where $z$ represents the propagation distance along the waveguide axis, $k_0$ is the wavenumber of light in the medium, the Laplacian, $\nabla^2$, is in the transverse $(x,y)$ plane, $\Delta n(x,y,z)$ is the refractive index change profile of the waveguides, and $n_0$ represents the refractive index of the ambient medium. The waveguides are fabricated to contain only one bound state, they can be approximately modeled using a nearest-neighbor tight-binding model:

$$H = -\sum_m (c_1 a_m^\dagger b_m + c_2 a_m^\dagger b_{m-1} + c.c. + i\gamma b_m^\dagger b_m / 2), \qquad (2)$$

where $a_m^\dagger$ ($a_m$) creates (destroys) a photon in site $m$ of the A sublattice, and similarly for $b_m^\dagger$ ($b_m$) in the B sublattice. The parameters $c_1$ and $c_2$ are hopping (coupling) constants between waveguides, and $\gamma$ is the loss rate. The hopping parameters are tuned by varying the spacing between sites, and the loss, $\gamma$, is controlled by varying the amplitude of the oscillating waveguides. In a waveguide geometry, this lattice model yields a discrete Schrodinger equation [30,31], with the distance along the waveguide axis, $z$, acting in place of time, $t$:

$$i\partial_z \begin{pmatrix} a_m \\ b_m \end{pmatrix} = -\begin{pmatrix} c_1 b_m + c_2 b_{m-1} \\ c_1 a_m + c_2 a_{m+1} - i\gamma b_m / 2 \end{pmatrix}. \qquad (3)$$

In the basis of periodic Bloch functions, the Hamiltonian reduces to a set of uncoupled 2x2 blocks, given by $H(k) = -(c_1 + c_2 e^{ikd})a_k^\dagger b_k + c.c. - i\gamma b_k^\dagger b_k/2$, where $k$ is the Bloch wave number. For $\gamma=0$, this is precisely the SSH model. This model, where the A and B sublattices are equivalent, possesses a symmetry which dictates that, for any $k$, the Bloch eigenstates (viewed as a pseudospin degree of freedom) lie on the equator of the Bloch sphere. The topological number in this model is the 'winding number:' namely, the number of loops made by a Bloch state around the equator of the Bloch sphere, as k passes through the first Brillouin zone [18,32–34].

Here, the winding number, w, yields $|w|=0$ or 1, in the cases $c_1>c_2$ and $c_1<c_2$, respectively; non-zero winding number is referred to as being 'topologically non-trivial'. Note the winding number is equivalent to the Zak phase [32,34], up to a factor of π. The bulk-edge correspondence principle [35] states that, at an interface between a topologically trivial and non-trivial crystal, a localized edge state forms. Typically, the topological nature of a system is probed via these edge states [5,13,16,26].

With the introduction of loss, the NHDM presents a different paradigm of topology [24]. Here, a winding number is associated with the number of times the Bloch eigenstates encircle the polar axis of the Bloch sphere, independent of the presence or absence of symmetry in the real parts of the A and B sublattice potentials. The z-axis is special due to the existence of a non-decaying dark state that lives entirely on the A sublattice, which is represented by the north pole of the Bloch sphere. Both the SSH and NHDM models feature a transition of the topological number at exactly $c_1=c_2$; in the case where A and B sublattice potentials are equal, the winding number of the NHDM precisely coincides with that of the SSH model. Remarkably, within the NHDM, the expected displacement, $\langle \Delta m \rangle$, achieved by a particle initialized on the non-decaying sublattice, is topologically quantized, with its value given by the winding number.

In our experiment, we first look at the case when light is injected (at $z=0$) with unit amplitude into the (non-lossy) A-site of unit cell $m=0$. This corresponds to an initial wavefunction of $\psi_0 = a_{m=0}^{(M)\dagger}|0\rangle$ (where $M$ represents the sublattice site excited). The mean displacement yields [24]:

$$\langle \Delta m \rangle = g \sum_m \int_0^\infty dz \, |\psi_m^{(B)}(z)|^2 \quad , \tag{4}$$

where $\psi^{(B)}_m(z)$ is the amplitude on site B of unit cell $m$ at longitudinal position $z$. Importantly, this quantity depends upon the wavefunction's history. It was shown in [24] that $\langle \Delta m \rangle$ is topological and represents the winding number if the A site is excited (M=A). Figure 2 shows $\langle \Delta m \rangle$ as a function of the deviation from the integer lattice $\Delta d = (d_1 - d_2)/2$, obtained from tight-binding simulations on a lattice of 400 unit cells. The coupling constant between two waveguides is an exponentially decreasing function of their separation, with a decay constant of

0.195/μm (which is sufficiently strong to neglect next-nearest-neighbor coupling). Therefore, the ratio $c_2/c_1$ exponentially increases with $\Delta d$, with $c_1 = c_2$ for $\Delta d=0$. The two curves (blue and red) correspond to initialization on the A and B sublattices, M=A and M=B, respectively.

The transition from $\langle \Delta m \rangle = 0$ to $\langle \Delta m \rangle = 1$ at $\Delta d=0$ is shown by the blue curve in Fig. 2, signifying the change of the winding number, and the system's switching between different topological phases. The red curve (for which $\langle \Delta m \rangle$ does not represent the winding number) remains zero for all $\Delta d$. The *z*-evolution of the wavefunction is plotted in the insets of Fig. 2 for $\Delta d=0$ (i.e., at the transition) and $\Delta d=2.8$μm (away from it), respectively. Interestingly, despite the loss in the system, the lifetime of the wavefunction diverges at $\Delta d=0$ (Fig. 2, right inset), however it is finite at $\Delta d=2.8 \mu m$ (Fig. 2, left inset). This is due to the presence of a 'dark state': a non-decaying eigenstate residing only on sites without loss [24,29], leading to a diverging variance, $\langle \Delta m^2 \rangle$, which accompanies the jump in $\langle \Delta m \rangle$ and signals the topological transition. This, along with the finite spatial and temporal extent of the simulation, accounts for the small degree of smoothness seen in the transition in Fig. 2.

To experimentally probe the transition, we fabricate such lattices in fused silica glass using the femtosecond direct laser writing technology [36] with fabrication parameters described in Ref. [29]. Each lattice consists of 34 waveguides and is 10cm long, along the waveguide axis. The waveguides are arranged with alternating distances $d_1$ and $d_2$ as depicted in Fig. 1. To introduce radiation losses (and so the non-Hermiticity), every second waveguide is fabricated with sinusoidal oscillations in the transverse direction as a function of z. We use an oscillation amplitude of A=3 μm and a period Z=1 mm, which results in a loss factor of $\gamma \approx 0.53$/cm. Every waveguide contains also some small intrinsic loss of $\gamma \approx 0.08$/cm arising from material absorption, which can be factored out as it is identical in every site. We implement 11 waveguide lattices with a lattice constant d=36 μm, with different distances $d_1$ and $d_2$ between adjacent sites, where $d_1+d_2=d$. In the most dimerized case we use $|\Delta d| = 4$ μm, which is decreased in each subsequent array in steps of 0.4 μm to $\Delta d = 0$, corresponding to equally spaced waveguides. When light is initially injected, there is some coupling loss because the input beam and the waveguide mode are not identical. However, this potential source of systematic error can be eliminated by extrapolating back to the origin to find the proper overall normalization factor.

In order to determine $\langle \Delta m \rangle$, we excite a single waveguide in the central unit cell (m=0) with laser light at λ=633nm, and employ fluorescence microscopy [36,37] to observe the light's propagation. The intensity distribution in the array is extracted from the fluorescence image. This allows determining the relative amplitude in each site at every position, *z*. From these data, we compute $\langle \Delta m \rangle$, as defined in Eq. (4). We choose the sample length sufficiently large such that, within our noise limit, practically no intensity is left within the waveguide array at the output facet, due to the radiation and absorption losses.

In Fig. 3, we plot $\langle \Delta m \rangle$ as a function of $\Delta d$. Here, we use the fact that two systems with the same modulus of Δd are realized in the same sample, which just has to be inverted during data analysis. One can see a clear transition from $\langle \Delta m \rangle \sim 0$ to $\langle \Delta m \rangle \sim 0.82$ when a straight waveguide in the central cell of the array (m=0) is excited (M=A). In the case where the B site is excited (M=B), $\langle \Delta m \rangle$ shows small fluctuations around zero, as expected from the tight binding limit investigated in Fig. 2.

Two examples of fluorescence images of the light intensity distribution in the lattices for $\Delta d = 0$ and $\Delta d = 2.8\ \mu m$ are shown in the insets to Fig. 3. Background noise is subtracted from the experimental data; however, there is inevitably some residual noise. The experimental curves in Fig. 3 (filled symbols) are compared to simulation results (unfilled symbols) that were obtained by performing continuum simulations using the beam propagation method.

In our experiment we attempt to see the topological transition in our non-Hermitian system via a bulk measurement. We provide the first observation of a dissipative topological transition. Namely, we see that the topological number defined in the NHDM undergoes a clear transition at the predicted value of Δd, the degree of dimerization. Of course, as expected for a continuous physical system, the transition shown in Fig. 3 is smooth, as opposed to the sharp transition predicted by the idealized tight-binding NHDM. The smoothening of the transition in the experiment and continuum simulations (Fig. 3), can be attributed to: (1) the tight-binding approximation not taking into account deviations from adiabaticity as the waveguides undergo oscillation to induce loss; (2) neglect of the overlap term, $\langle \psi | \phi \rangle$, between neighboring waveguide modes $|\psi\rangle$ and $|\phi\rangle$; (3) assuming that the loss in the B waveguides yields exponential decay, while in reality there can be resonances with continuum leaky modes [38]). The differences between the experimental results and continuum simulations can arise due to (1) background radiation noise (especially at large *z* where the signal to noise ratio goes to zero); and/or (2) deviations from a linear relationship between fluorescence and the light intensity guided within the waveguides. These also account for the reduced experimental value of *<Δm>*=0.82 in the non-trivial phase instead of the ideal value 1. Despite these, the setting captures very well the qualitative behavior of the topological transition. It even gives accurate quantitative results for the point of transition, as well as the fact that there is no transition at all when the system is excited at the B site.

Another prediction of Ref. [24] is that the lifetime of the wavefunction should diverge at the transition, here corresponding to Δd>0, monotonically decreasing for increasing |Δd|. To investigate this, we plot the integrated power in the final quarter of the sample as a function of Δd (Fig. 4). We show only *Δd>0*, as the inverted structures used to realize negative values of Δd yield identical results. Clearly, the lifetime is maximal at the transition, although – as in any experiment – lifetime does not diverge. The lack of a divergence at *Δd=0* arises simply because of intrinsic absorption loss associated with the waveguides. This calculation was performed without an overall renormalization of the wavefunction (as was required for Fig. 3, as discussed earlier).

An important point bears discussion here, namely, that under inversion, the waveguide structure with a given dimerization Δd maps to the waveguide array structure with dimerization -Δd. This raises the question of how these two structures can have different winding numbers, w – while being exactly the same. The answer is that, as in the SSH model, there is indeed an ambiguity if we simply consider the lattice bulk. In the SSH model, symmetry is broken by the edge: if the edge starts with a long bond, this defines the unit cell, and means an edge state exists and the winding number is 1. If it starts with a short bond, no edge state exists and the winding number is zero. In our NHDM system, the edges play no role, hence the ambiguity is resolved by the (arbitrary) choice of unit cell labeling. This arbitrariness represents a necessary feature arising from the fact that we do not probe edge properties of this system, but rely solely on bulk information. More specifically, specific phases cannot be labeled as topologically trivial or non-trivial; what is important is only the topological distinction between phases as signaled by their differing winding numbers (see discussion in [18]). Therefore, the transition should be viewed as a probe of the difference in the winding numbers of the two phases in the NHDM.

In conclusion, we have presented the first experimental observation of a topological transition in a non-Hermitian system; we do this using bulk measurements only. We used a non-Hermitian bipartite system: a dimer with loss on every other site [24]. Our experiments show that the features of the idealized tight-binding model of Ref. [24] indeed carry over to experiments. In other words, the non-Hermiticity allows the recovery of the topological features. Our experiments are carried out in a photonic system, but many of the concepts are generally applicable to any continuous system where non-Hermiticity can be readily realized. This work opens up alternative schemes for characterizing the topology of systems where edge behavior is difficult or impossible to probe. Indeed, loss is an inherent feature of many systems that may exhibit topological effects (such as coupled quantum dots [24], optical lattices [18], and driven-dissipative exciton polariton condensates [39]). This non-Hermiticity may act as a surprising probe of their potentially rich topological phases.

[1] D. J. Thouless, M. Kohmoto, M. P. Nightingale, and M. den Nijs, Phys. Rev. Lett. **49**, 405 (1982).
[2] K. v. Klitzing, G. Dorda, and M. Pepper, Phys. Rev. Lett. **45**, 494 (1980).
[3] C. L. Kane and E. J. Mele, Phys. Rev. Lett. **95**, 226801 (2005).
[4] B. A. Bernevig, T. L. Hughes, and S.-C. Zhang, Science **314**, 1757 (2006).
[5] M. König et al., Science **318**, 766 (2007).
[6] F. D. M. Haldane and S. Raghu, Phys. Rev. Lett. **100**, 013904 (2008).
[7] Z. Wang, Y. D. Chong, J. D. Joannopoulos, and M. Soljačić, Phys. Rev. Lett. **100**, 013905 (2008).
[8] Z. Wang, Y. Chong, J. D. Joannopoulos, and M. Soljacic, Nature **461**, 772 (2009).
[9] R. O. Umucalılar and I. Carusotto, Phys. Rev. A **84**, 043804 (2011).
[10] M. Hafezi, E. A. Demler, M. D. Lukin, and J. M. Taylor, Nat. Phys. **7**, 907 (2011).
[11] K. Fang, Z. Yu, and S. Fan, Nat. Photonics **6**, 782 (2012).
[12] A. B. Khanikaev et al., Nat. Mater. **12**, 233 (2013).
[13] M. C. Rechtsman, et. al, Nature **496**, 196 (2013).
[14] M. Hafezi et al., Nat. Photonics **7**, 1001 (2013).
[15] Y. E. Kraus et al., Phys. Rev. Lett. **109**, 106402 (2012).
[16] T. Kitagawa et al., Nat. Commun. **3**, 882 (2012).
[17] A. V. Poshakinskiy et al., Phys. Rev. Lett. **112**, 107403 (2014).
[18] M. Atala et al., Nat. Phys. **9**, 795 (2013).
[19] D. A. Abanin, T. Kitagawa, I. Bloch, and E. Demler, Phys. Rev. Lett. **110**, 165304 (2013).
[20] X.-J. Liu, K. T. Law, T. K. Ng, and P. A. Lee, Phys. Rev. Lett. **111**, 120402 (2013).
[21] A. Dauphin and N. Goldman, Phys. Rev. Lett. **111**, 135302 (2013).

[22]  G. Jotzu et al., ArXiv14067874 Cond-Mat Physics, quant-Ph (2014).
[23]  T. Ozawa and I. Carusotto, Phys. Rev. Lett. **112**, 133902 (2014).
[24]  M. S. Rudner and L. S. Levitov, Phys. Rev. Lett. **102**, 065703 (2009).
[25]  S. Diehl, E. Rico, M. A. Baranov, and P. Zoller, Nat. Phys. **7**, 971 (2011).
[26]  D. Hsieh et al., Nature **452**, 970 (2008).
[27]  W. P. Su, J. R. Schrieffer, and A. J. Heeger, Phys. Rev. Lett. **42**, 1698 (1979).
[28]  N. Malkova et al., Opt. Lett. **34**, 1633 (2009).
[29]  T. Eichelkraut et al., Nat. Commun. **4**, (2013).
[30]  F. Lederer et al., Phys. Rep. **463**, 1 (2008).
[31]  A. L. JONES, J. Opt. Soc. Am. **55**, 261 (1965).
[32]  B. A. Bernevig and T. L. Hughes, *Topological Insulators and Topological Superconductors* (Princeton University Press, Princeton, 2013).
[33]  P. Delplace, D. Ullmo, and G. Montambaux, Phys. Rev. B **84**, 195452 (2011).
[34]  J. Zak, Phys. Rev. Lett. **62**, 2747 (1989).
[35]  S. Ryu and Y. Hatsugai, Phys. E Low-Dimens. Syst. Nanostructures **22**, 679 (2004).
[36]  A. Szameit and S. Nolte, J. Phys. B At. Mol. Opt. Phys. **43**, 163001 (2010).
[37]  A. Szameit et al., Appl. Phys. Lett. **90**, 241113 (2007).
[38]  O. Peleg et al., Phys. Rev. A **80**, 041801 (2009).
[39]  T. Jacqmin et al., Phys. Rev. Lett. **112**, (2014).

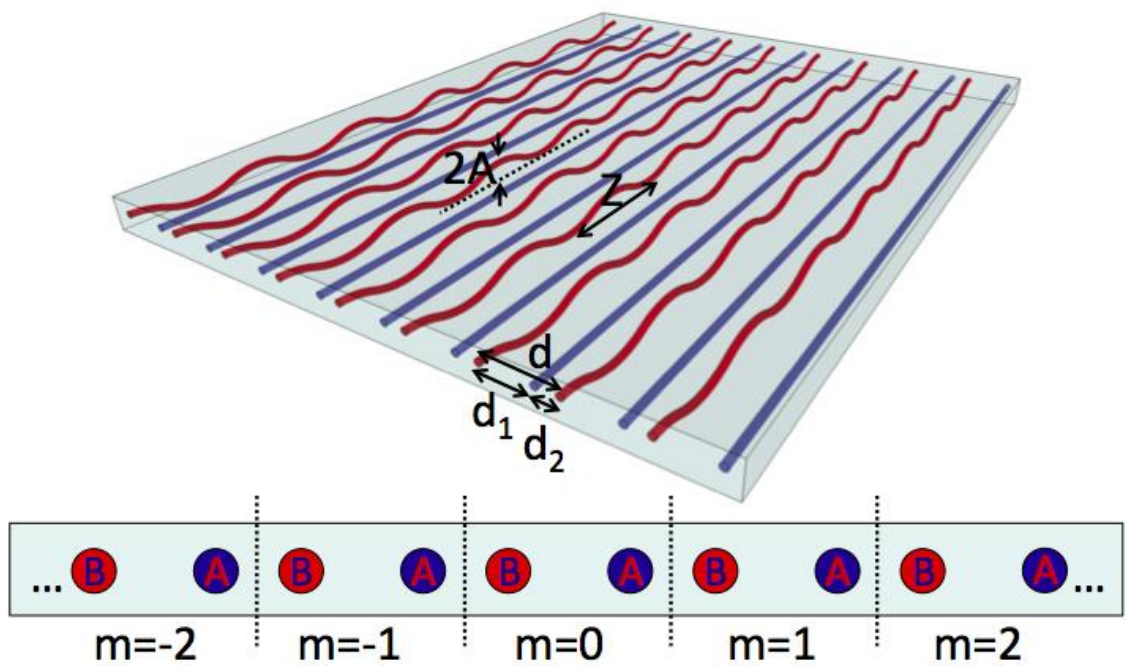

Figure 1. (color online) Schematic diagram of the waveguide array structure. The B sites oscillate in the vertical direction with a period $Z = 1$mm, causing bending/radiation losses to continuum modes.

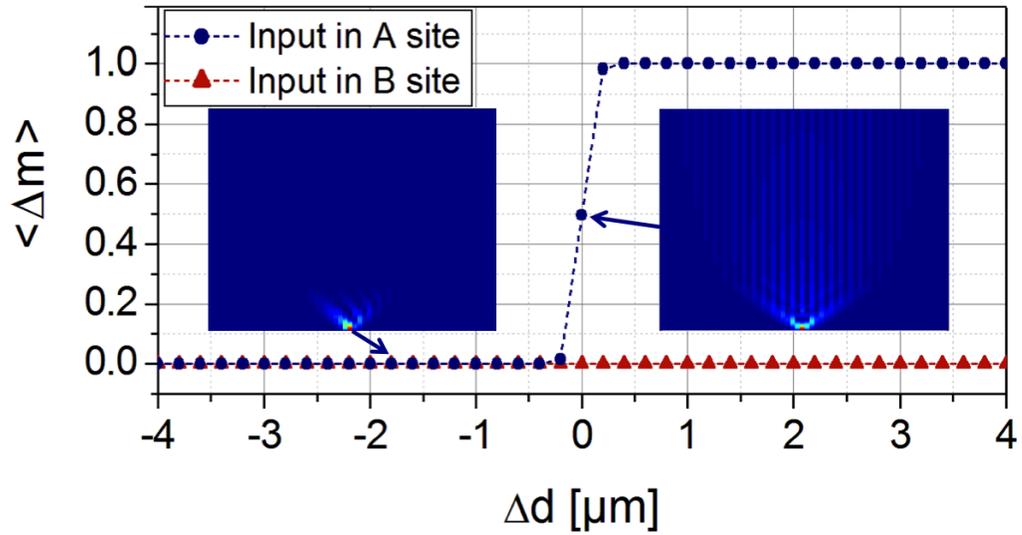

Figure 2. (color online) Mean displacement <Δm> plotted as a function of Δd, the structural deviation from the integer lattice. The blue curve depicts the result when light is input in the A site and the red curve corresponds to input at the B site. Insets show the absolute value of the wavefunction as a function of z (ascending) in the waveguide array (note the presence of the long-lived 'dark state' on the right inset).

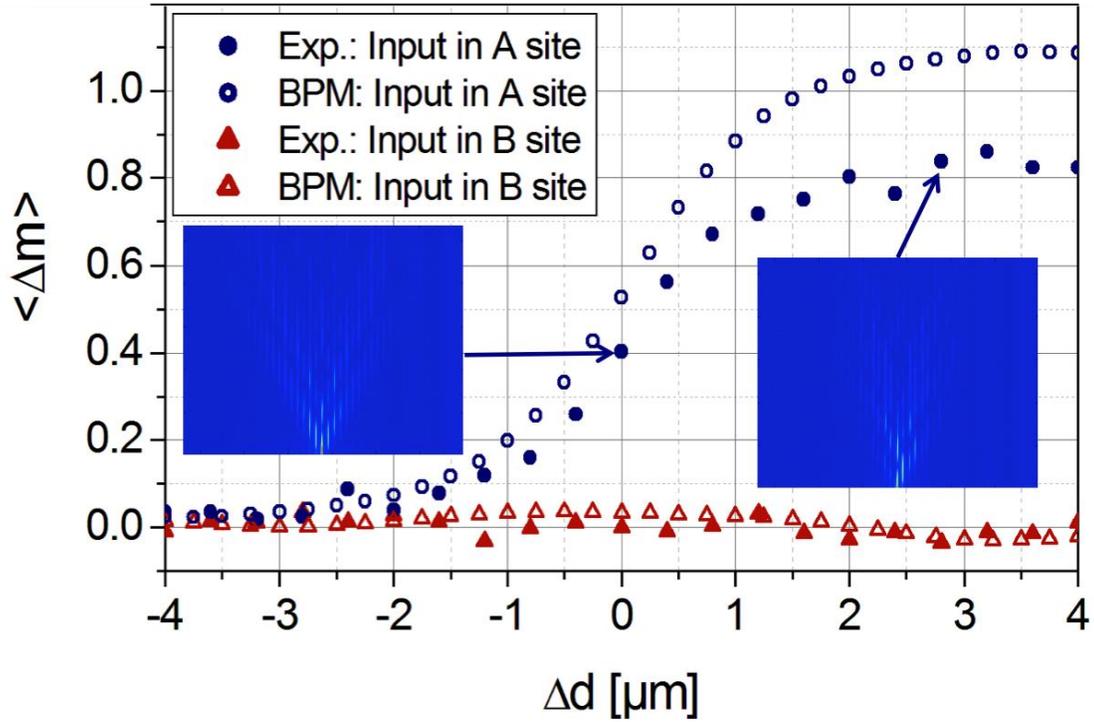

Figure 3. (color online) Experimental results showing the mean displacement <Δm> vs. structural deviation Δd. The blue curves are for light input on the non-lossy sites (M=A), and the red curves are for light input on the lossy sites (M=B). The insets show the evolution of the absolute value of the wavefunction in the experiment at two different values of Δd, 0μm and 2.8μm.

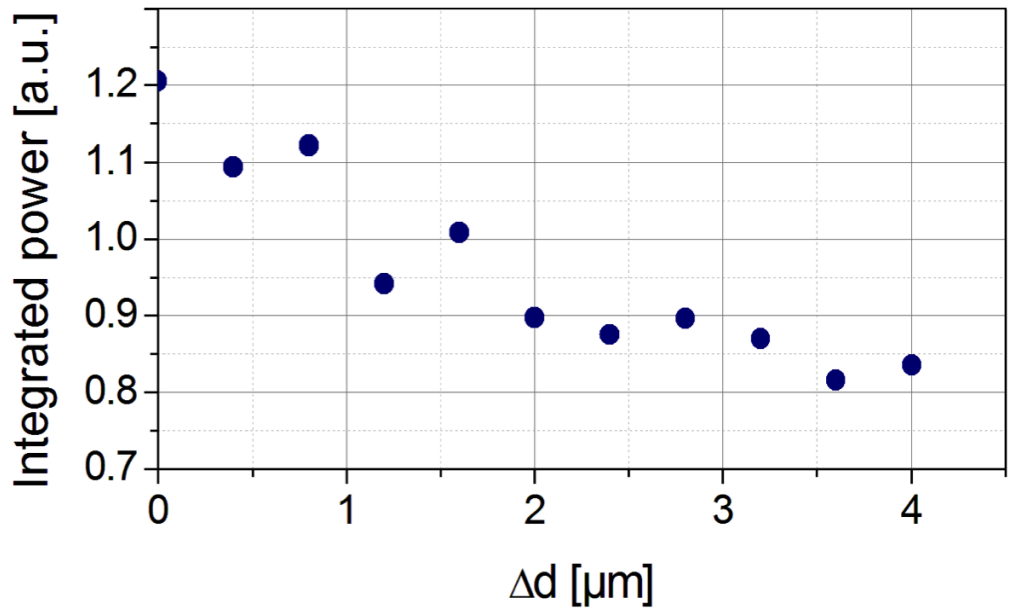

Figure 4. Experimental results showing the total integrated power in all waveguides in the last 2.5cm of propagation through the array, as a function of the structural deviation from the integer lattice Δd.